\documentstyle[12pt]{article}
\def\ni{\noindent}

\def\be{\begin{equation}}
\def\ee{\end{equation}}
\def\bea{\begin{eqnarray}}
\def\eea{\end{eqnarray}}
\def\ni{\noindent}
\begin{document}
\ni
{\large\bf C-axis resistivity and high-T$_c$ superconductivity}\\

\ni
N. Kumar,* T.P. Pareek** and A.M. Jayannavar**\\

\ni* Raman Research Institute, Bangalore 560080, India\\
**Institute of Physics, Bhubaneswar 751005, India
\vspace*{0.2cm}

\begin{abstract}
Recently we had proposed a mechanism for the normal-state C-axis
resistivity of the high-T$_c$ layered cuprates that involved {\em
blocking} of the single-particle tunneling between the weakly coupled
planes by strong intra-planar electron-electron scattering. This gave
a C-axis resistivity that tracks the ab-plane T-linear resistivity, as
observed in the high-temperature limit.  In this work this mechanism is
examined further for its implication for the ground-state energy and 
superconductivity of the layered cuprates. It is now argued that,
unlike the single-particle tunneling, the tunneling of a boson-like pair
between the planes prepared in the BCS-type coherent trial state remains
{\em unblocked} inasmuch as the latter is by construction an eigenstate
of the pair annihilation operator. The resulting pair-delocalization
along the C-axis offers energetically a comparative advantage to the
paired-up trial state, and, thus stabilizes superconductivity.  In this
scheme the strongly correlated  nature of the  layered
system enters only through the {\em blocking} effect, namely that a given
electron is effectively repeatedly {\em monitored} (intra-planarly
scattered) by the other electrons acting as an environment, on a
time-scale shorter than the inter-planar tunneling time. Possible
relationship to  other inter-layer pairing mechanisms proposed by
several workers in the field is also briefly discussed.
\end{abstract}
{PACS numbers: 74.20, 74.72.-h,03.65.-w, 34.30+h}
\baselineskip=24pt
\par The two electronic-structural features, now become central to a proper
understanding of the normal-state resistivity as well as the
high-temperature superconductivity of the highly anisotropic marginal
metals, namely the layered cuprates, are the strong electron-electron
correlation and their effectively low (two) dimensionality.$^{1,2}$  We
have, thus, the oxygen-hole doped CuO$_2$ planes representing the
strongly correlated electronic system, while the weak inter-planar
tunneling through the thick spacer layers of the reservoir oxides, e.g.,
SrO, Bi$_2$O$_3$, etc., gives the near two-dimensionality. In our recent
work,$^{2,3}$ it was shown that these two features, namely the strong
intra-planar electron-electron scattering and the weak interplanar
tunneling, can give rise to a C-axis resistivity that tracks the T-linear
metal-like ab-plane resistivity in the high-temperature limit, with an
insulator-like upturn at low enough temperatures.  Also, the metal-like
C-axis resistivity (\(\rho_c(T)\)) can have a magnitude not bounded by
Mott's maximum metallic resistivity. These results are in qualitative
agreement with the measured $\rho_c(T)$ on high quality single-crystal
samples that reflect, presumably, their intrinsic transport behaviour.$^5$
Furthermore, the C-axis transport is found to be  necessarily incoherent
as indeed supported by observations.$^{2,6}$  This mechanism for incoherent
C-axis transport was also proposed independently by Leggett,$^6$ and has
now been followed up by a number of workers in the field.$^7$  The physics
underlying our proposed mechanism is that of the {\em blocking} of the
weak interplanar tunneling by the relatively strong intraplanar inelastic
scattering. This is, of course, a particular case of the celebrated
Quantum Zeno effect $-$ namely, the suppression of transition between two
weakly coupled Hilbert subspaces due to strong intra-subspace coupling to
environment.$^8$  Thus, in the present case the two neighbouring CuO$_2$
planes (e.g., of the bilayer), coupled weakly through a small inter-planar
tunneling matrix-element, constitute the two electronic subspaces, and the
strong intra-planar scattering of a given electron by the other electrons
represents the intra-subspace environmental coupling. We will now examine
this {\em blocking} effect further for its implication for the
ground-state energy of and for the superconductive electron-pairing in
these layered strongly correlated systems.  Our main finding is that the
strong-intra-planar electron-electron scattering does indeed, at zero
temperature,  {\em block} the single-electron inter-layer tunneling but
{\em not} the tunneling of (the time-reversed) electron pairs. The
resulting inter-planar pair-delocalization energetically favours the
pairing globally and hence stabilizes superconductivity. The calculation
is done for a simple bilayer model. The present work is much in the spirit
of, and complements the work of, Chakravarty, Subdo, Anderson and
Strong$^9$ and that of Kumar,$^{10}$ all based on the idea of confinement by
orthogonality catastrophe.$^{11}$    We also discuss in this context how the present
mechanism differs essentially from the several other pairing mechanisms
that involve inter-layer tunneling.

Let us first consider the possible {\em blocking} of the single-electron
inter-planar tunneling due to intra-planar scattering at zero-temperature.
Now, in the high-temperature limit the in-plane inelastic scattering can
be viewed as a stochastic field acting on a given electron attempting to
tunnel out-of-plane. This general picture is well known and well
supported, experimentally as well theoretically, in the context of
decoherence.$^{11,12}$  The problem becomes rather subtle at low (zero)
temperature, and is best probed in the present context by calculating the
change $\Delta E_o$ of the ground-state energy $E_o$ of a weakly coupled
bilayer as function of the strength ($\lambda$) of the inplane
electron-electron scattering, maintaining, of course, the system in the
normal state, i.e., without breaking spontaneously any global symmetry,
such as the one 
responsible for superconductivity.  {\em Blocking} effect is expected to
reduce the change
$\Delta E_o$ as $\lambda$ is increased. This is readily concluded by using the
Hellmann-Feynaman {\em charging} technique involving in the present case
an integration with respect to the intra-planar tunneling matrix element
($-t_\perp$) as the variable coupling parameter. 

The Hamiltonian for a bilayer of the weakly  coupled planes, labelled A and
B, can be written as
\be
H = H_A + H_B + H_{AB}({\rm single\,\,\,particle})\,,
\ee
where the intra-planar Hamiltonians $H_A$ and $H_B$ describe the two
interacting electron subsystems of the isolated planes A and B, and
\be
H_{AB} ({\rm single\,\,\,particle}) = - \eta t_\perp \Sigma_{{\bf k}\sigma}
(a_{{\bf k}\sigma}^\dagger b_{{\bf k}\sigma} + b_{{\bf k}\sigma}^\dagger
a_{{\bf k}\sigma})\,,
\ee
with $t_\perp = t_\perp^* > 0$, the tunneling Hamiltonian with the
creation/annihilation fermionic operators \(a_{{\bf k}\sigma}^{\dagger}
(b_{{\bf k}\sigma}^\dagger) / a_{{\bf k}\sigma} (b_{{\bf k}\sigma})\)
referring to 
the planes A(B).  Here tunneling is taken to conserve the in-plane
wavevector {\bf k} and the spin projection $\sigma$.  The tunneling
matrix element $-t_\perp$ is taken to be small in a sense to be made
precise later.  The dimensionless parameter $\eta$ is to be set equal to 1
at the end.  The exact ground-state energy $E_o(\eta)$ of the bilayer
varies with $\eta$ parametrically according to the Hellmann-Feynman
theorem as
\be
\frac{\partial E_o (\eta)}{\partial \eta}  =  < \eta | H_{AB} | \eta >
\end{equation}
giving
\be 
\Delta E_o \equiv E_o(1) - E_o(0) = -2t_\perp \int_o^1 d\eta \Sigma_{\bf k} < \eta
|a_{\bf k}^\dagger b_{\bf k} | \eta >\,,
\ee
where $|\eta>$ denotes the exact bilayer-ground
state for a given value of the parameter $\eta$.  Here we have dropped the
spin projection label $\sigma$, and the wavevector summation includes
summation over $\sigma$.

Expressing the equal-time correlation $<\eta| a_{\bf k}^\dagger b_{\bf
k}|\eta >$ in terms 
of the imaginary part of the retarded Green function \(G_{AB}^R ({\bf k},w)
(\equiv G_\perp^R ({\bf k},w))\), we get
\be
\Delta E_o = \frac{2t_\perp}{\pi} \Sigma_{\bf k} \int_o^{\infty} d\omega \int_o^1 d\eta {\rm Im}
G_{\perp}^R ({\bf k},
w)
\ee
Now, the exact retarded Green function $G_\perp^R$ for the correlated
metallic planes A and B coupled by the weak tunneling $-\eta t_\perp$ is
clearly not known.  We can, however, adopt the following viewpoint. In the
absence of the inter-planar tunneling, the correlated electron planes A
and B can be well modelled by the semi-phenomenological Marginal
Fermi-Liquid (MFL)$^{13}$ which is known to be consistent with the T-linear
ab-plane resistivity. The corresponding retarded in-plane ($\parallel$)
Green functions $G_{AA}^R 
(k,w) = G_{BB}^R ({\bf k},w) \equiv G_{\parallel}^R ({\bf k},w)$ is then given by
\be
G_\parallel^R ({\bf k},w) = \frac{1}{w - \epsilon_{\bf k} - \Sigma^R ({\bf
k}, w)}
\ee
with
\begin{eqnarray*} 
{\rm Re} \Sigma^R ({\bf k}, w) &=& \lambda w  ln (w/w_c)\\
{\rm Im} \Sigma^R ({\bf k}, w) &=& -\lambda \frac{\pi}{2} w
\end{eqnarray*}
with $w_c > w > 0$.
Henceforth we will drop the superscript $R$.

\ni
Now, for sufficiently small $t_\perp$, one can assume the
electron-electron scattering to take place on a time scale $\ll$ the
tunneling time $h/t_\perp$, and, therefore, ignore the vertex corrections
to the inter-planar tunneling.  We can then at once write down from the
Dyson equation for the retarded inter-planar ($\perp$) Green function
$G_\perp$: 
\be
G_\perp ({\bf k},w) = \frac{\eta t_\perp (G_\parallel ({\bf k},w))^2}{1 - \eta^2 t_\perp^2 (G_\parallel ({\bf k}, w))^2}
\ee
Now, substituting from Eqs. (6) and (7) into Eq. (5). and performing the
{\bf k}-integration with a constant 2D-density of states $n_o$, we get
\be
\delta e = \int_o^\infty dw [I(W, t_\perp, w) - I(-W, t_\perp, w) + I(W,
-t_\perp, w) - I(-W, -t_\perp, w)],
\ee
where dimensionless energy change
\be
\delta e = \frac{\Delta E_o}{\left (\frac{4n_o t_\parallel^2}{\pi}\right )}\,,
\ee
\newpage
with
\bea
I(W, t_\perp, w) &=& - t_\perp {\rm arctan} \left [\frac{W + t_\perp - w +
\lambda w ln (\frac{w}{w_c})}{\frac{\pi}{2} \lambda w}\right ] - \left(W - w +
\lambda w ln(\frac{w}{w_c})\right) \nonumber\\
&&{\rm arctan} \left[\frac{t_\perp w \lambda \frac{\pi}{2}}{\left\{\left(W
+ t_\perp - w + \lambda w ln (\frac{w}{w_c})\right ) \times \left(W - w +
\lambda w ln(\frac{w}{w_c})\right ) + \left ( \lambda w \frac{\pi}{2}\right
)^2\right \}}\right ]\nonumber\\
&& + \frac{\pi}{4} \lambda w ln \frac{\left[\left(W + t_\perp - w + \lambda
w ln (\frac{w}{w_c})\right )^2 + \left (\lambda w \frac{\pi}{2}\right
)^2\right ]}{\left[\left (W - w + \lambda
w ln (\frac{w}{w_c})\right )^2 + \left (\lambda w \frac{\pi}{2}\right
)^2\right ]}
\eea 
\noindent where $W$ is the two dimensional bandwidth.
\par In Fig.1, we have plotted the energy gain (reduction $-\delta e$ in the
ground-state energy), due to delocalization by inter-planar single-particle
tunneling (-$t_\perp$), against the intra-planar electron interaction
strength $\lambda$.  It is readily seen that the energy gain is
a sharply decreasing function of the intra-plane interaction strength
$\lambda$.  This clearly demonstrates the effective {\em blocking} of the
single-particle inter-planar tunneling by intraplanar scattering $-$ the
Quantum Zeno effect $-$ as 
anticipated on physical grounds.

Thusly encouraged, we now address the rather subtle question as to how
this blocking of the single-particle tunneling becomes ineffective against
the tunneling of (time-reversed) bosonic pairs. In order to clearly appreciate this
point, let us consider the two electronic-subsystems forming the
bilayer to be prepared in a BCS-like trial many-body state $\psi$ (in the absence
of tunneling).  Thus we have for the decoupled bilayer,
\bea
|\psi>\,\,\,\,\,\,\,\,\,\,\,\,\,\, &=& \prod_{{\bf k},q} \left(u_{\bf k} + v_{\bf k} a_{{\bf
k}\uparrow}^\dagger  a_{-{\bf k}\downarrow}^\dagger \right ) \left (u_{\bf
q} + v_{\bf q} b_{{\bf q}\uparrow}^\dagger  b_{-{\bf
q}\downarrow}^\dagger\right ) | 0 >\nonumber\\
\propto e^{\phi(\alpha^\dagger + \beta^\dagger)} | 0 > &\equiv& | A > | B >\,,
\eea
where $\alpha^{\dagger} {(\alpha)},  \beta^\dagger (\beta)$  are the pair creation
(annihilation) operators for the two planes A and B of the bilayer. Here
\begin{eqnarray}
\phi \alpha^\dagger &=& \Sigma_{\bf k} \left(\frac{v_{\bf k}}{u_{\bf
k}}\right ) a_{{\bf k}\uparrow}^\dagger  a_{-{\bf k}\downarrow}^\dagger\\
\phi \beta^\dagger &=& \Sigma_{\bf k} \left(\frac{v_{\bf k}}{u_{\bf
k}}\right ) b_{{\bf k}\uparrow}^\dagger b_{-{\bf k}\downarrow}^\dagger\,,
\end{eqnarray}
where we can take as usual the operators $\alpha^\prime$s and $\beta^\prime$s to be
bosonic to a good approximation. Thus, the unnormalized trial function $|\psi>$ is a
coherent state, i.e., a phased superposition of states with different
number of pairs, with
$|\phi|^2$ representing eventually the mean bosonic pair-occupation number for each
of the planes A and B.

However, these trial coherent states $|A>$ and $|B>$ are certainly not the
ground states for the isolated 2D-electronic subsystems A and B, with
(repulsive) electron-electron interaction in general.  The crucial
observation, however,  is that the coherent states $|A>$ and $|B>$ are,
respectively, eigenstates of the pair annihilation operators $\alpha$
and $\beta$.  If, therefore, we now introduce a pair-tunneling
($-t_\perp^\prime$) term in the
Hamiltonian, $H_{AB} ({\rm pair}) = -t_\perp^\prime (\alpha^\dagger \beta
+ \beta^\dagger \alpha)$ (with $t_\perp^\prime  \neq t_\perp$ in
general), we at once verify that
\be
<\psi | H_{AB} ({\rm pair}) |\psi> = -2t_\perp^\prime |\phi|^2\,.
\ee
This implies adiabatic transfer of the bosonic pairs between the two planes A and
B of the now coupled bilayer prepared in the coherent state $|\psi>$.  This
inter-planar pair delocalization in turn stabilizes the trial state
$|\psi>$ energetically. Thus, given that the single-particle tunneling
($-t_\perp$) is blocked effectively while the pair-tunneling
($-t_\perp^\prime$) is not as suggested by the  above, we can expect
$t_\perp^\prime \gg t_\perp$ and  the coherent state $|\psi>$
to be stabilized relative to the normal state. This is the central point of
the pairing mechanism proposed in this work.

Once this is accepted, the energetic stabilization of such a BCS-like
paired up state due to the dominance of the pair tunneling over the
single-particle tunneling can be readily treated within a mean-field
approximation. For this, consider the reduced Hamiltonian that should
suffice for describing the low-energy phenomena:
\[
H_{red} ({\rm bilayer}) = H_A + H_B + H_{AB} ({\rm single\,\,\,particle}) + H_{AB}({\rm pair})
\] 
with
\begin{eqnarray}
H_A &=& \Sigma_{{\bf k}\sigma} \epsilon_{\bf k} a_{{\bf k}\sigma}^{\dagger} a_{{\bf k}\sigma} + U_{\rm
eff} \Sigma_{\bf k k^\prime} a_{{\bf k}\uparrow}^\dagger  a_{-{\bf k}\downarrow}^\dagger  a_{-{\bf
k^\prime}\downarrow} a_{{\bf k^\prime}\uparrow}\nonumber\\
H_B &=& \Sigma_{{\bf k}\sigma} \epsilon_{\bf k} b_{{\bf k}\sigma}^{\dagger} b_{{\bf k}\sigma}  + U_{\rm
eff}
     \Sigma_{\bf k k^\prime} b_{{\bf k}\uparrow}^\dagger  b_{-{\bf k}\downarrow}^\dagger  b_{-{\bf
k^\prime}\downarrow} b_{{\bf k^\prime}\uparrow}\nonumber\\
H_{AB} ({\rm single}) &=& -t_\perp \Sigma_{\bf k} (b_{{\bf k}\sigma}^\dagger
a_{{\bf k}\sigma} + h.c.)\nonumber\\
H_{AB} ({\rm pair}) &=& -t_\perp^\prime \Sigma_{\bf k k^\prime} (b_{{\bf
k}\uparrow}^\dagger 
b_{-{\bf k}\downarrow}^\dagger  a_{-{\bf k^\prime}\downarrow} a_{{\bf k^\prime}\uparrow}
+ h.c.), 
\end{eqnarray}
where $t_\perp^\prime \gg t_\perp > $0, and $U_{\rm eff}$ can even
be moderately repulsive ($U_{\rm eff}$ $>$ 0) as  considered by Zecchina,$^{14}$
except for the retention of the single-particle tunneling here. The latter
enables us to treat the effect of the degree of blocking of
single-particle tunneling explicitly. Note that the reduced Hamiltonian
maintains the condition of pairing involving electrons in time-reversed states.

Consider first the case of $U_{\rm eff}$ negative (attractive).
Introducing the anomalous averages in the spirit of the mean-field 
approximation (MFA), we get 
\bea
H_{MFA} &=& \Sigma_{{\bf k}\sigma} \epsilon_{{\bf k}\sigma} a_{{\bf
k}\sigma}^\dagger a_{{\bf k}\sigma} + \Sigma_{\bf k} \epsilon_{{\bf
k}\sigma} b_{{\bf k}\sigma}^\dagger b_{{\bf k}\sigma} \nonumber \\
&+& \Delta V \Sigma_{\bf k} \left (a_{-{\bf k}\downarrow} a_{{\bf k}\uparrow} +
a_{{\bf k}\uparrow}^\dagger a_{-{\bf k}\downarrow}^\dagger \right )\nonumber\\ 
&+& \Delta V \Sigma_{\bf k} \left (b_{-{\bf k}\downarrow} b_{{\bf k}\uparrow} +
b_{{\bf k}\uparrow}^\dagger b_{-{\bf k}\downarrow}^\dagger \right )\nonumber\\ 
&-& t_\perp \Sigma_{\bf k} \left (b_{{\bf k}\sigma}^\dagger a_{{\bf
k}\sigma} +
a_{{\bf k}\sigma}^\dagger b_{{\bf k}\sigma}\right )\,,
\eea
where the s-wave gap parameter
\[\Delta = \Sigma_{\bf k^\prime}  \left <a_{-{\bf k^\prime}\downarrow} a_{{\bf
k^\prime}\uparrow}\right > =
\Sigma_{\bf k^\prime}\left < b_{-{\bf k^\prime}\downarrow} b_{{\bf k^\prime}\uparrow} \right >\]\\ 
and \[V = \left(\frac{U_{\rm eff} -t_\perp^\prime}{2}\right).\]
After straightforward diagonalization, the self-consistent gap equation
for $\Delta$ turns out to be
\bea
\Delta = &-& \frac{1}{2} \Sigma_{{\bf k}} \frac{\Delta V(1-2f(\epsilon_{\bf k}-
t_\perp))}{2\sqrt{\Delta^2V^2 + (\epsilon_{\bf k}- t_\perp)^2}}\nonumber\\
 &-& \frac{1}{2} \Sigma_{\bf k} \frac{\Delta V(1-2f(\epsilon_{\bf k}+
t_\perp))}{2\sqrt{\Delta^2V^2 + (\epsilon_{\bf k}+  t_\perp)^2}}\,,
\eea
where as usual $f$ is the fermi function, \(f(\epsilon) =
\frac{1}{(e^{\epsilon/k_BT}+1)}.\) The corresponding equation for the 
critical temperature $T_c$ is then (for $U_{\rm eff} - t_\perp^\prime < 0$)
\be
1 = -\left(\frac{U_{\rm eff} - t_\perp^\prime}{2}\right) (\frac{1}{2}) \Sigma_{\bf k}
\left[\frac{1}{2{\left(\epsilon_{\bf k} - t_{\perp}\right)}}
\tanh \left (\frac{\epsilon_{\bf k} - t_\perp}{k_BT_c}  \right ) +
\frac{1}{2 (\epsilon_{\bf k} + t_\perp)} 
\tanh \left (\frac{\epsilon_{\bf k} + t_\perp}{k_BT_c}  \right )  \right ]
\ee
This reduces to the usual expression in the limit $t_\perp \rightarrow$0
(i.e., total blocking of the single electron tunneling). It is also
readily seen from the sub-linear x-dependence of $\tanh x$ that incomplete
suppression of the single-particle blocking ($t_\perp\neq 0$) leads to a reduction in
$T_c$. Thus we recover our claim that the blocking of the single-particle
tunneling relative to the pair-tunneling stabilizes the paired-up
superconducting state.

Some remarks are in order at this point. For an attractive $U_{\rm eff}$,
the present pair-tunneling mechanism may well be viewed as an
amplification of the (s-wave) superconductive pairing pre-existing in the
isolated planes. (This, of course, remains true also for the case when the
isolated planes support d-wave pairing, arising from the spin-fluctuation
mechanis, say.)  We would, however, like to emphasize here that our
present mechanism provides for a 
global stabilization of the condensed state even when the pairing
potential for the individual pairs ($U_{\rm eff}$) is repulsive, but , of
course, sufficiently small and the isolated planes are not superconducting
on their own.  Thus, for a short-ranged repulsive  potential  our
inter-layer mechanism based on the Zeno-effect can stabilize a coherent
condensate, albeit of  d-wave pairs.  (The case of a strongly negative
$U_{\rm eff}$ supporting a repulsive bound state lying above the top of
the band is quite different. It can give high lying d-wave pairs that may
get stabilized coherently through interlayer pair tunneling mechanism.) 
Indeed, the present mechanism involves global stabilization, and  cannot
be reduced to a pairing potential arising, say, from virtual exchange of
some excitations. 

It will be apt now to view the present mechanism  in relation to
others involving pairing by inter-layer coupling as reported in the
literature. First, let us note that the important role of interlayer
coupling is strongly suggested by the experimental fact that the bilayers
(or the intimate layers in general) are a necessary minimum for the
occurrence of superconductivity in these layered cuprates.$^{15}$  There
is also a definite evidence for the Josephson coupling between these
layers.$^{16}$ Furthermore, it has become increasingly clear that a simple
one-band two-dimensional Hubbard model does not support high-T$_c$
superconductor.  All this has prompted several workers to propose and
treat models involving interlayer coupling, or indeed, plain-chain or
chain-chain coupling.$^2$  These could be introduced phenomenologically
through the macroscopic Josephson coupling of the order parameters,$^{17}$
or microscopically through the introduction of interplanar pair
tunneling,$^{14}$ or through interplanar pair-pair interactions.$^{18}$
The essential contrasting feature of our present work is our demonstration
of the blocking of the single-particle tunneling (our equation 8), and our
argument for the unblocking of the pair tunneling (our equation 14). 
Our work is closest in spirit that of Chakravarty and Anderson$^9$
which it complements.  It is, however, different from the earlier
inter-layer pair mechanism of Wheatley, Hsu and Anderson$^{19}$ which
involves spin charge separation and 
exchange of spinons mediating the interlayer tunneling of a pair of the
otherwise confined physical electrons of opposite spins.  It is, however,
quite likely that the non-Fermi liquid feature involved in this exotic
model is in the ultimate analysis a yet another and novel route to
realizing the Quantum Zeno 
effect in the extreme total confinement by the orthogonality catastrophe.

In conclusion, we have extended the mechanism proposed by us earlier for
the C-axis resistivity, involving the blocking of the interplanar
single-particle tunneling by the intra-planar scattering, to low
temperatures to possibly
explain the High-T$_c$ superconductivity of the layered cuprates. We have
given an argument based on coherence and supported by simple analysis
that, in-contrast to the single-particle tunneling, the tunneling of the
bosonic pairs remains unblocked and thus stabilizes the superconducting state.
In this scheme, the strongly correlated nature of the two-dimensional
layers enters only through this single-particle blocking effect. The
present mechanism admits d-wave pairing without recourse to pairing mechanisms
such as that of the spin-fluctuation theory.$^{20}$

\newpage
{\bf Figure Caption}
\newline
{\bf Fig.1} Dimensionless energy gain due to interplanar single particle tunneling
as function of intraplanar interaction parameter $\lambda$. The decreasing
energy gain with increasing $\lambda$ demonstrates the blocking effect.
The plot is for the choice of parameters $(t_{\parallel}/t_{\perp})$=20.0 and
$(\omega_{c}/t_\perp)$=100.
\end{document}